\documentclass[aps,prb,groupeaddress]{revtex4}

\usepackage{amsmath} 
\usepackage{amssymb} 
\usepackage{amsfonts}
\usepackage[dvips]{graphicx} 
\usepackage[]{epsfig}

\begin{document}
\bibliographystyle{apsrev} 

\title{Ion-specificity in $\alpha$-helical folding kinetics} 
\author{Yann von Hansen} 
\author{Immanuel Kalcher} 
\author{Joachim Dzubiella} 
\thanks{To whom correspondence should be addressed. E-mail: jdzubiel@ph.tum.de}
\affiliation{Physics Department T37, Technical University  Munich, 85748 Garching, Germany.}

\begin{abstract}

The influence of the salts KCl, NaCl, and NaI at molar concentrations on the $\alpha$-helical folding 
kinetics of the alanine-based  oligopeptide Ace-AEAAAKEAAAKA-Nme is investigated by means of (explicit-water)
molecular dynamics simulations and a diffusional analysis.  The mean first passage times for folding and
unfolding are found to be highly salt-specific. In particular, the folding times increase about one order 
of magnitude for the sodium salts. The drastic slowing down can be traced back to long-lived, compact  
configurations of the partially folded peptide, in which sodium ions are tightly bound by several carbonyl and carboxylate
groups.  This multiple trapping is found to  lead to a non-exponential residence time distribution of the cations in 
the first solvation shell of the peptide. The analysis of  $\alpha$-helical folding in the framework of diffusion in a reduced (one-dimensional) free energy  landscape  further shows that the salt not only specifically modifies equilibrium properties, but also induces kinetic barriers due to individual ion binding.  In the sodium salts, for instance,  the peptide's configurational mobility (or "diffusivity") can decrease about one order of magnitude.  This study demonstrates the highly specific action of ions and highlights the intimate coupling of intramolecular friction and solvent effects in protein folding. 

\end{abstract}

\maketitle 
\section{Introduction}

The complex problem of protein folding is typically interpreted in terms of a diffusive search through an effective, low-dimensional free energy landscape, where most of the countless degrees of freedom of the system have been integrated out.~\cite{Bryngelson1989,Camacho1993,Socci1996,Dill1997,Hummer2000,Best2006} 
In this picture the effective diffusivity (or friction) arising from intrapeptide and peptide-solvent interactions as well as from orthogonal degrees of freedom varies considerably along the reaction coordinate (RC).~\cite{Best2010, Hinczewski2010} In general, the internal friction constitutes a major contribution to the friction, such that solvent viscous drag and solvent-peptide interactions are not the only mechanisms that govern protein kinetics.~\cite{Pabit2004,Cellmer2008}  Internal and solvent-induced friction processes may even be intimately connected  as was demonstrated for the loop formation rate of unfolded peptides, where the strongly denaturing salt guanidine hydrochloride modifies internal friction by specific binding mechanisms.~\cite{Buscaglia2006, Moglich2006}  

The investigation of  the salt-specific action on proteins in general ('Hofmeister effects') has a long history,~\cite{Baldwin1996} but the underlying mechanisms are still under exploration. In a recent series of experiments, for example, it has been shown that even the  simple cations sodium (Na$^{+}$) and potassium (K$^{+}$) exhibit considerably different behavior in the interaction with protein surfaces, where Na$^+$ is  favored over K$^+$.~\cite{Vrbka2006,Uejio2008,Aziz2008} One apparent reason is the stronger attraction of sodium to acidic (anionic) surface groups, in particular to carbonyl and side chain carboxylate groups. While these static properties have received much attention lately,~\cite{Fedorov2009} not much is known about their consequences to biomolecular kinetics.  Experimental hints have been given in studies of Na$^{+}$- and K$^{+}$-specific polyglutamic acid aggregation kinetics,~\cite{Colaco2008} folding kinetics of halophilic ('salt-loving' and very acidic) proteins~\cite{Madern2000,Bandyopadhyay2007} or DNA.~\cite{Gray2008}  However, a detailed molecular understanding of the ion-specific action on biomolecular folding and assembly kinetics is still lacking. 

In particular for large concentrations ($c\gtrsim 1$~M) salt effects are pronounced, 
highly sequence and salt-type specific, and typically lead to changes in protein solubility, stability,
 and/or denaturation that result in the so-called Hofmeister series for the precipitation of proteins.~\cite{Baldwin1996}  Although an order of magnitude higher than at typical physiological conditions ($c \sim 0.2$~M), 
large salt concentrations play a central biochemical role in the broad field of protein 
crystallization,~\cite{Dumetz2007} in food industry as fermentation additives,~\cite{Dyer1951} 
and for the function and stability of biotechnologically interesting halophilic (salt-loving) enzymes.~\cite{Lanyi1974} 
Additionally, the  study of protein structures in solvents with high salinity solvents is instructive as salt-specific effects are amplified 
and, important from a computational perspective,  can be  sampled more efficiently in molecular dynamics 
(MD) simulations, a popular tool nowadays for the theoretical study of protein folding, 
function, and stability.~\cite{Karplus2002}

An ubiquitous and fundamental secondary structure element of proteins is the $\alpha$-helix, which is 
stabilized by $(i,i+4)$ backbone hydrogen bonds involving four amino acids (aa) per turn. 
The majority of short ($\lesssim$ 20 aa) isolated helices derived from proteins are unstable in solution, unless 
specific side-chain interactions stabilize them. Among those it has been demonstrated that alanine-based peptides have the strongest intrinsic helix propensity.~\cite{Marqusee1987,Marqusee1989,Spek1999, Chakrabartty1994, Scholtz1991}  A very instructive model peptide was experimentally introduced by 
Marqusee and Baldwin \cite{Marqusee1987}, showed that  alanine-based oligopeptides with an A(EAAAK)$_n$A pattern display a high $\alpha$-helix propensity, which is probably stabilized by  Glu$^{-}$ (E) and Lys$^{+}$ (K) salt bridges along the folded peptide. Marqusee and Baldwin  also found that the simple salt NaCl has a destabilizing effect on the $\alpha$-helical 
configurations of these peptides.

Indeed, in a recent explicit-water molecular dynamics (MD) computer simulation\cite{Dzubiella2008} the structural behavior 
of the oligopeptide Ace-AEAAAKEAAAKA-Nme,  in the following named the 'EK' peptide,  has been explored in detail,
and the stabilizing and destabilizing mechanisms in various highly concentrated aqueous electrolyte solutions 
have been identified. Among those it has been found that sodium (Na$^+$) destabilizes the helical structure stronger
than potassium (K$^+$); it was also previously recognized that the reason for the destabilization of the salt bridges and the secondary-structure forming hydrogen lies in the higher affinity of Na$^+$ to carboxylate and carbonyl groups. In a recent follow-up paper the folding kinetics of the EK-peptide in pure aqueous solution was investigated and interpreted in terms of diffusion in a reduced (one-dimensional) free energy landscape involving a local coordinate-dependent diffusivity.~\cite{Hinczewski2010} 

The aim of this work is to extend these previous studies on the 'EK'-peptide to the investigation of  the specific effects of the salts KCl, NaCl, and NaI  on peptide $\alpha$-helical {\it folding and unfolding  kinetics}. In particular, we  explore how salts alter the mean folding and unfolding times and look for possible molecular reasons. Strongly salt-specific effects are found in the mean folding times, which can be attributed to the binding of individual ions to multiple, anionic peptide groups inducing transient cross-links between peptide fragments.
In consequence, not only equilibrium distributions of configurations are modified, but also new, slow time scales in the peptide's configurational mobility arise due to enhanced internal friction.  
Salt effects are thus reflected in both modified free energy landscape and local changes of the effective diffusivity. 
Our study demonstrates the highly specific action on protein folding kinetics by the individual binding of ions, and, more generally, exemplifies the intimate coupling between solvent and intrapeptide friction effects in protein folding. 
We believe that these mechanisms could be of  general importance and transferable to a variety of biomolecular and polyelectrolyte systems. 

\section{Methods}

\subsection{MD  simulations}

Our MD simulations are performed using the parallel module {\it sander.MPI} in the simulation package Amber9.0 with the ff03 force field for the peptides and the rigid and nonpolarizable TIP3P  water for the 
solvent.~\cite{Amber}  All simulated systems are maintained at a fixed pressure of $P=1$ bar and a temperature $T=300$~K by coupling to a Berendsen barostat and Langevin thermostat,~\cite{Amber} respectively. The cubic and periodically repeated simulation box of edge length $L~\simeq 36$~\AA~ includes  approximately 1500 water molecules. Electrostatic interactions are  calculated by particle mesh Ewald summation and all real-space interactions (electrostatic and  van der Waals) have a cut-off of 9~\AA. The peptide is generated using the {\it tleap} tool in the Amber package.~\cite{Amber}

We investigate the helical folding and unfolding behavior of a 12 amino acids long peptide with the acetyl (Ace) and amine (Nme) capped sequence Ace-AEAAAKEAAAKA-Nme. This peptide can  form three $\alpha$-helical turns in the fully folded state,  where Glu2 and Lys6, and Glu7 and Lys11, are potentially able to form a salt bridge, respectively.\cite{Dzubiella2008}  The influence on $\alpha$-helix folding kinetics of a large concentration of $ \simeq 3.6\pm$0.1 M of the simple  monovalent salts NaCl, KCl, and NaI is investigated. We have simulated the system without salt for $\simeq 1.35~\mu$s and including salt for $\simeq 2~\mu$s for each salt type. We note here that the free energy along the RC q (see below) 
derived from replica-exchange MD trajectories at $T=300$~K for the salt-free system gives very good agreement to the brute force approach~\cite{Hinczewski2010} indicating a decent statistical sampling by our trajectories.

The considered salt concentrations result from $90$ ion pairs in the simulation box. Cations and anions are modeled as nonpolarizable Lennard-Jones spheres with charge and interaction parameters as supplied by Dang,~\cite{Dang1995} as the default Amber parameters are known to be faulty.~\cite{Joung2008} The Dang parameters show reasonable bulk thermodynamic properties in SPC/E water even for high concentrations.~\cite{Kalcher2009} Comparative calculations (unpublished) in TIP3P water show only small differences in hydration structure, and no qualitative difference in the binding to peptide 
groups.~\cite{Dzubiella2010}.
The parameters used are summarized in previous work on the equilibrium structure of the EK peptide in salt at a different temperature.~\cite{Dzubiella2008} 
We are aware of the weaknesses of ionic MD force fields for quantitative predictions in biomolecular simulations; this issue has already been discussed in our previous study, where however a reasonable description of helicity and destabilization with NaCl was observed when compared to experiments.
Since the destabilization seems to be overemphasized for NaCl, we do not claim to be quantitative in  our work, but focus on the discussion of the main effects and qualitative trends with the addition of salt; we believe these to be insensitive to the particular force field and relevant for a variety of experimental observations.
 
\subsection{Helicity and reaction coordinate}

Trajectory analysis is  performed using the {\it ptraj}  tool  in the Amber9.0 package.~\cite{Amber}
The helicity, i.e., the $\alpha$-helical fraction, is identified using the DSSP method by Kabsch and Sander.~\cite{Kabsch1983}
We focus on one RC $Q$, which is defined as the root mean square distance from a fully helical reference structure (i.e., with helicity equals to one), averaged over all atoms of the peptide and which thus measures the deviation from the 'native' state;  this quantity was previously found to act as an adequate dynamic RC in the salt-free case.~\cite{Hinczewski2010} 
We note that different choices of the reference structure were tested  and resulted only in small differences in the energy landscapes, i.e., local 
variations on the order of fractions of the thermal energy $k_\text{B}T$; see the Supporting Information (SI) for additional details.
Similarly, the RC-trajectories only marginally change when varying the reference structure and therefore yield the same kinetic behavior; examples of such trajectories are also found in the SI. 
Trajectories are recorded with a resolution of 20~ps, giving a total of  $\sim$ 67500 data points for  the simulations without salt and roughly 100000 data points for the runs including salt.
For convenience, we define a rescaled RC by $q = (Q - Q^\text{min})/(Q^\text{max}-Q^\text{min})$ such that the minimal and maximal values of the  data points, denoted as $Q^\text{min}$ and $Q^\text{max}$, are projected on the RC values $q=0$ and $q=1$, respectively.
The absolute minimum and maximum RC values are similar for all systems and are $Q^\text{min}\simeq 1.0$~\AA~  and $Q^\text{max}\simeq 8.0$~\AA.
 
\subsection{Bulk shear viscosities}
In order to get a more complete picture of the solvent properties we calculate bulk shear viscosities for NaCl, KCl, and NaI at the relevant concentrations in TIP3P water. 
We employ the same ionic force fields as above, but perform the simulations with the GROMACS~4.0~\cite{vanderSpoel2005,Lindahl2001} package due to the implemented viscosity calculation methods. 
In these simulations, the periodically repeated box has an edge length of $L\simeq4~\text{nm}$, with a total number of $N_\text{w}=1910$ water molecules and $N_\text{i}=135$ ion pairs. 
For the pure water simulation we use $N_\text{w}=2180$ water molecules. 
After $NPT$-equilibration we proceed with an $NVT$-production run of $50~\text{ns}$.
We compare two approaches to calculate the viscosity: firstly, we employ the Green-Kubo (GK) formula~\cite{Hess2002,Zwanzig1965}
\begin{equation}
\eta~=~\frac{V}{k_{\text{B}}T}\,\int^{\infty}_{0}\left<P_{xz}(t_0)P_{xz}(t_0+t)\right>_{t_0}\,\text{d}t,
\end{equation}
involving the off-diagonal components of the pressure tensor only.
We obtained averaged viscosities over a correlation time of $5$ to $20~\text{ps}$. The latter choice reflects the fact that the viscosity converges rapidly, but exhibits large statistical errors for long correlation times.~\cite{Chen2009}

Secondly, we perform  a non-equilibrium perturbation method.~\cite{Hess2002} In this scheme, an external force is applied in the $NVT$-simulation with the periodic acceleration profile
\begin{equation}
a_x(z)~=~A\cos(kz),
\end{equation}
with $k=\frac{2\pi}{L}$, $L$ being the edge length of the box. The amplitude $A$ should be chosen small enough in order not to drive the system out of the linear response regime and at the same time large enough to get good statistics. For a more detailed discussion we refer to previous work~\cite{Hess2002} and set the amplitude to $A=0.02~\text{nm}~\text{ps}^{-2}$. We then obtain the viscosity by calculating the average velocity profile of all particles.~\cite{Hess2002}

With the GK formula we find values of $\eta_0=0.31$ $\pm0.01$ 10$^{-3}$~kg/(m$\cdot$s) for pure TIP3P water, corroborating with previous studies,~\cite{Gonzalez2010} and $\eta=$ 0.58, 0.74, and 0.6$\pm0.01$ 10$^{-3}$~kg/(m$\cdot$s)for the KCl, NaCl, and NaI solutions at a concentration of 3.6~mol/l, respectively. 
The periodic perturbation method yields the same results within a $5\%$ error range. 
Compared to experimental values~\cite{Robinson} the MD simulation considerably overemphasizes the increase of the viscosity at this elevated salt concentration; indeed, the viscosity was experimentally found to increase by only roughly 5\% for KCl and 30-40 \% for NaCl and NaI compared to pure water. 
This failure in describing the correct bulk viscosities of the electrolyte solutions must be attributed to inaccuracies in the force field. 
Note that the value for pure TIP3P water already considerably deviates from the experimental value (0.893 $\cdot 10^{-3}$~kg/(m$\cdot$s) at 298.15~K)~\cite{Harris2004} by more than a factor of two.

\subsection{Free energy and diffusivity profiles} 

We assume that the stochastic time evolution of the probability $\Psi(q,t)$ of finding a
configuration with RC value $q$ at time $t$ is described by the one-dimensional
Fokker-Planck (FP) equation~\cite{Hanggi1990}
\begin{equation}
\label{eq:FokkerPlanck}
\frac{\partial}{\partial t} \Psi(q,t) = \frac{\partial}{\partial q}
 D(q)   \text{e}^{-\beta F(q)}  \frac{\partial}{\partial q}  \Psi(q,t) \text{e}^{\beta F(q)},
\end{equation}
where $D(q)$ is the (in general $q$-dependent) diffusivity, $\beta\equiv1/(k_\text{B}T)$, and $F(q) = - k_\text{B}T\ln \langle \Psi(q) \rangle $ is the free energy obtained by Boltzmann-inversion of the time-averaged probability distribution $\langle \Psi(q)\rangle$.
We employ a recently introduced method to estimate the diffusivity profile
that takes advantage of the relation between average transition times between different points along the RC, the free energy landscape $F(q)$, and the diffusivity profile $D(q)$.~\cite{Hinczewski2010}  The round-trip time defined as 
\begin{equation}
\label{eq:round-trip}
\begin{split}
\tau_\text{RT}(q, q_\text{t}) \equiv\text{sign}(q-q_\text{t}) [ \tau_\text{fp}(q, q_\text{t}) + \tau_\text{fp}(q_\text{t}, q)],\end{split}
\end{equation}
specifies the average time needed for an excursion starting at $q$, reaching $q_\text{t}$ at least once, and returning to $q$ for the first time; mean first passage times for transitions starting at $q$ and reaching $q_\text{t} $ are denoted by $\tau_\text{fp}(q, q_\text{t})$.
For a diffusive process described by the FP equation~(\ref{eq:FokkerPlanck}) the round-trip time is given by the integral
\begin{equation}
\label{eq:round-trip_FP}
\tau_\text{RT}(q, q_\text{t}) =
\int_{q_\text{t}}^q\text{d}q^\prime\;\frac{1
}{D(q^\prime)\langle \Psi(q^\prime)\rangle}, 
\end{equation}
which can be derived from the expressions for the mean first passage time given before~\cite{Hinczewski2010}. 
Though Eq.~(\ref{eq:round-trip_FP}) can in principle be inverted to obtain the diffusivity profile $D(q)$ from the slope of the round-trip time curves, we choose a complementary analysis method within the present work to avoid artifacts due to insufficient statistical sampling.
The FP approach assumes an underlying Markovian process and --- as is easily seen from Eq.~(\ref{eq:round-trip_FP}) --- round-trip time curves (as a function of $q$) for different target points $q_\text{t}$ therefore only differ by a constant
\begin{equation}
\label{eq:round-trip_addditivity}
\tau_\text{RT}(q, q_\text{t})=\tau_\text{RT}(q, q_\text{t}^{\prime})+\tau_\text{RT}(q_\text{t}^{\prime}, q_\text{t}).
\end{equation}
The assumption of Markovian behavior however generally breaks down at short times and for unsuitable RCs, i.e., RCs that do not single out the transition state; see also previous literature on that topic.~\cite{Hinczewski2010}

In our analysis of the simulation time series $q(t)$ we discretize the RC in $N=50$ intervals centered around $q^{(i)}=(2i-1)/100,\;i\in\{1,2,\dots,50\}$; mean first passage times between all possible pairs of bins are extracted from simulation data and converted into round-trip times using Eq.~(\ref{eq:round-trip}). To simplify the analysis, we assume a flat diffusivity within each of the following regions: (i) values of the RC $q < q_{2/3}$ corresponding to an almost perfectly folded helix, (ii) one or two unfolded helical turns equivalent to $q_{2/3}\leq q< q_{1/3}$, and (iii) mostly unfolded states characterized by RC values $q\geq q_{1/3}$, where the indices 1/3 and 2/3 denote the average helicity at these $q$-values. The values of the diffusivity in those three regions are used as fit parameters in Eq.~(\ref{eq:round-trip_FP}) to best reproduce the round-trip time curves obtained from simulation data; the integral in Eq.~(\ref{eq:round-trip_FP}) is computed numerically by linear interpolation of $\{\langle\Psi(q^{(i)})\rangle\}_{i=1}^{50}$.
Best fits to the round-trip data for different target points $q_\text{t}$ allow to determine the diffusivity (including an error estimate) for each of the three regions.
Alternatively, the diffusivity profile can also be obtained by fitting to the average round-trip time
\begin{equation}
\label{eq:round-trip_avg}
\bar{\tau}_\text{RT}(q^{(j)})\equiv\frac{1}{N}\sum_{i=1}^N\tau_\text{RT}(q^{(j)},q^{(i)}), 
\end{equation}
which is less affected by statistical noise and which according to Eq.~(\ref{eq:round-trip_addditivity}) is just shifted vertically w.r.t. the round-trip curve $\tau_\text{RT}(q,q_\text{t})$ for a specific target position $q_\text{t}$.
The results of both fitting procedures are shown and discussed in Sec.~\ref{sec:Results_kinetics}, additional information concerning the fitting is provided in the SI.

\section{Results and Discussion}

\subsection{Equilibrium free energy landscapes}

The RC time series $q(t)$ of all investigated systems is shown in Fig.~1:  
numerous folding-unfolding transitions on different time scales are discernible.
Already at a first glance longer time scales (and fewer unfolding-folding transitions) seem to be involved in the systems with the sodium salts, which 
will be quantified and discussed in more detail later. 
Free energy profiles $F(q)$ of all investigated systems are extracted from the trajectories $q(t)$ and plotted in Fig.~2 together with the (time-averaged) helicity resolved along the RC $q$.
In the salt-free case (black curve) in Fig.~2~(a), three minima are clearly visible: one at $q_1\simeq 0.1$, a second at $q_2\simeq 0.32$, and a 
third, shallow one at $q_3=0.58$. 
Judging from the helicity vs. $q$, plotted in Fig.~2~(b), these minima seem to correspond mainly to 1) a full helix with three turns, 2) a partially unfolded helix with two neighboring turns, and 3) one helical turn as well as fully unfolded states. 
Representative simulation snapshots taken at corresponding $q_i$-values ($i=1,2,3$) confirm this view and are also shown in  Fig.~2~(b). 
We note that the value of $q_1$, which in absolute units corresponds to $Q_1\simeq 1.6$~\AA, deviates from 0 (the reference state) due to thermal fluctuations.

For KCl, the free energy $F(q)$ is slightly shifted to favor unfolded states at  larger $q$-values ($q_3\simeq 0.67$),  while the main features remain the same as without salt. 
The total helicity decreases slightly from 62\% to 55\%. 
For the sodium salts, NaCl and NaI, the fully folded state at $q_1$ becomes metastable, and the distribution is strongly shifted to the partially and fully unfolded states. 
In particular, the third minimum at $q_3$ deepens and broadens. 
Also in the systems with salt the minima mostly match with partially folded $\alpha$-helical states as can be judged from the helicity vs. $q$ curve. 
All total helicities and positions of the minima in $F(q)$ are  summarized in Tab.~I. 

In the NaCl and NaI salts the total helicity significantly decreases to about 39\% and 34\%, respectively. The main causes
have been discovered before~\cite{Dzubiella2008} and are twofold: 
firstly, specific Na$^+$-binding to the glutamic acid side chain carboxylates interferes with E$^-$-K$^+$ salt bridging, 
and secondly,  the binding of cations to backbone carbonyls perturbs secondary-structure forming hydrogen 
bonds.~\cite{Dzubiella2008} In comparison to  Na$^+$, the specific binding of K$^+$ is weak and KCl 
thus behaves less destabilizing. We note that quantitatively these effects are force field dependent while
the binding trends have been corroborated in many experiments and various simulation studies.~\cite{Vrbka2006,Uejio2008,Aziz2008} For completeness, we plot radial distribution functions (rdfs) between the cations and the oxygen atoms from the 
backbone carbonyls and side chain carboxylates in Fig.~3. The dominance of sodium over potassium binding 
to the anionic peptide groups is clearly observable.

\begin{table}[h]
  \caption{Total helicity of the investigated systems and distinct positions in the free energy landscape shown in Fig.~2: locations of local minima are denoted by $q_i,\,i=1,2,3$, and values of the RC, where the average helicity is 2/3 or 1/3, by $q_{2/3}$ and $q_{1/3}$ respectively.}
\begin{center}
\begin{tabular}{| c | c | c | c | c| c |c|c|}
\hline
       system      &  helicity & $q_1$   &     $q_2$&     $q_3$  &$q_{2/3}$ &$q_{1/3}$\\  
   \hline
       no salt  & 0.62 & 0.11     &    0.31  & 0.57& 0.34& 0.56  \\
       KCl      &  0.55 & 0.10     &    0.32  & 0.67 & 0.35& 0.65 \\
       NaCl    &  0.39 & 0.08      &    0.28  & 0.59  & 0.30& 0.52\\
       NaI      &  0.34 & 0.09       &    0.31  & 0.65   & 0.32& 0.53\\
  \hline
\end{tabular}
\label{tab1}
\end{center}
\end{table}

\subsection{Long-lived structures and specific ion binding}

Previous work on the EK-peptide stability in salt solution~\cite{Dzubiella2008} revealed that due to Na$^+$ binding {\it long-lived} loop-forming  configurations occupy the region of intermediate to large $q$-values, in addition to one-turn and random coil states also present without salt. 
In these looped configurations a single Na$^+$-ion is collectively trapped by a few peptide backbone carbonyls and side chain carboxylates. 
This leads to a partial wrapping of the oligomeric backbone around the ion.   From a superficial inspection of our trajectories we find that these long-lived configurations involving tightly bound Na$^+$-ions can be stable on a $\simeq$~1-10 ns time scale. Representative simulation snapshots which are dynamically selected such that they exist for more than 2~ns  are shown in Fig.~4: in panel (a) a sodium ion is bound and wrapped by the central part of the peptide, while the terminal parts exhibit partial helical turns.  The bound sodium 'locks' this structure on times up to several nanoseconds.  Similar long-lived situations are found, where two neighboring turns are correctly folded, while the rest of the peptide forms a loop around the cation, see Fig.~4~(b). Often also the glutamic acid side chain is involved as displayed in Figs.~4~(c) and (d) where the cation not only binds to a backbone carbonyl but also to 
the headgroup of the E$^-$ side chain. In all these situations the peptide is relatively compact with one or two turns correctly
folded such that the relevant $q$-region of these states is between $0.25 \lesssim q \lesssim 0.65$, where the helicity is mostly between 
$2/3$ and $1/3$ (cf. Fig.~2~(b)). Configurations of this type have not or only rarely been found involving anions or a K$^+$-ion, so that  their existence must be attributed to the relatively strong binding of the Na$^+$ ion.  

The rigorous life-time analysis of the occurring configurations involving trapped ions is 
difficult to perform due to the variety and complexity of the somehow amorphous structures.  We therefore resort to the analysis of cation binding times. In the inset to Fig.~3, we plot the binding time distribution $P_b(t)$ of the cations K$^+$ and Na$^+$ in the first solvation shell  of carboxylates and carbonyls (defined by the location of the first minimum in the cation rdfs in Fig.~3). While the distribution decays exponentially for K$^+$ with a time constant of about 50 ps, we find a much slower, nonexponential  behavior for Na$^+$ which can be best fitted by a stretched exponential of the form $\propto\exp(-t/\tau)^\beta$ with $\tau\simeq 50$~ps
and $\beta=0.55$.  This indicates that long binding times on the order of nanoseconds are indeed possible and corroborate the 
existence of 'trapped' unfolded configurations in which peptide parts tightly wrap around the cation.  

We note that it is indeed well established that systems with multiple trapping or other manifestations of disorder can lead to anomalous kinetics.\cite{Metzler2000} 
In our investigated systems, the appearance and magnitude of trapping is controlled by the nature of the ions. The observed 'stretched' exponentials resemble the slow relaxation in glass-forming liquids.~\cite{Barrat} 
Anomalous kinetics in peptide and protein dynamics have been indeed observed in  simulations and experiments for certain dynamic variables or RCs,~\cite{Kou2004,Neusius2008, Sangha2009} and obviously question the general validity of diffusive approaches to predict long time dynamics in protein folding.
While this complex issue is still awaiting resolution we proceed in this work with the interpretation of helix folding in the framework of simple diffusion; since the average folding / unfolding times ($\gtrsim 10~\text{ns}$) are typically much larger then the ion binding times ($\sim 50~\text{ps}$), we expect the long-term dynamics to be adequately  described in terms of memoryless diffusion in a free energy landscape.

\subsection{Folding kinetics and diffusional analysis}
\label{sec:Results_kinetics}

Let us first analyze mean folding and unfolding times. In Fig.~5.~(a) we plot the mean first passage time for folding
from $q>q_1$ to $q_1$ given by $\tau_\text{f}(q,q_1)\equiv\tau_\text{fp}(q,q_1)$; the salt specific values of $q_1$ are found in Tab.~\ref{tab1}.  Without salt the typical folding time
is about $\tau_\text{f}\simeq 20-30$~ns in the region $q\gtrsim 0.3$ before it quickly drops down to 0 for $q$-values
closely approaching $q_1$.  In KCl $\tau_\text{f}(q,q_1)$ increases by a factor of about 2, while the sodium salts
lead to a considerable slowing down of folding by one order of magnitude. The unfolding times $\tau_\text{uf}(q,q_3)$ ($q<q_3$), 
cf. Fig.~5~(b), show a bit less variation between the salts. Without salt the typical unfolding time is about 30-40 ns while
it may rise by a factor of 2-3 in NaCl or KCl. Note that for KCl the unfolding is considerably slower than in the salt-free case 
although the free energy landscape is very similar. 

To get a grasp on the folding kinetics involving fewer helical turns we have also analyzed 
$\tau_\text{f}(q,q2)$ for $q>q_2$ (folding by 1 or 2 turns to the 2-turn state) and $\tau_\text{uf}(q,q_2)$ for $q<q_2$ (unfolding by 1 turn to the 2-turn state), which are shown in Fig.~6~(a) and (b), respectively. The folding times decrease by a factor of about 2 when compared to the $q\rightarrow q_1$ folding, while the trends with salt remain the same, in particular a one order of magnitude slower folding in the sodium salts. Unfolding times $\tau_\text{uf}(q,q_2)$ are relatively small and found between 3-7 ns. The variation between the salts is again less pronounced for unfolding when compared to the folding times and show a different ordering. 

Note that the mean first passage times shown in Figs.~5 and 6 are in parts subject to substantial noise due to insufficient statistical sampling: for example, clear deviations from a monotonously increasing function (expected for diffusive dynamics) are observed in the KCl-folding times shown in Fig.5~(a); this irregularity has its origin in an exceptionally long-lived (relative to the full trajectory duration)  state of a specific peptide configuration in the KCl-time series in Fig.~1 within the time range $200~\text{ns}\lesssim t\lesssim 400~\text{ns}$. Without going too much into detail, this specific configuration is characterized by two helical turns and the Ace-cap being buried between the hydrophobic side chains of one Lys and one Ala side chain.  This configuration (not involving bound ions) is found rather frequently in all trajectories but with typically much shorter life times.   

Let us now turn to the interpretation of the mean folding times by the free energy $F(q)$ and diffusivity profiles $D(q)$. 
For statistical reasons leading to the above mentioned anomalies, we have resolved the $D(q)$ profile only by three $q$-regions
distinguishing between the three states: 1) mainly full-helix, 2) one and two-turn states, and 3) mainly unfolded states. 
We assume that these states are separated by the $q$-values, where the average helicity is 2/3 and 1/3 in Fig.~2 (the salt-specific values of $q_{2/3}$ and $q_{1/3}$ being summarized in Tab. I). Although the resolution of $D(q)$ is small we emphasize that it can almost quantitatively reproduce the folding and unfolding times plotted in Figs.~5 and 6. Mean first passage times for folding and unfolding predicted by the FP equation involving salt-specific free energies and diffusivities are shown in the SI. 

Fig.~7 shows diffusivity profiles obtained by fitting to the average round-trip time defined in Eq.~(\ref{eq:round-trip_avg}) (solid lines) and the $D(q)$-estimates resulting from fits to round-trip times of specific targets $q_\text{t}$ (symbols with error bars): first of all, we observe that they are not flat, a feature  discussed previously for this peptide in the salt-free case.~\cite{Hinczewski2010} 
The inhomogeneities reflect variations of the multi-dimensional configurational mobility of the peptide projected onto the one-dimensional RC $q$.
Note also the fact that both analysis methods yield estimates, which coincide within error bars (the only exception being KCl for $q>q_{1/3}$), by this clearly validating our approach.
After including the salt a few  significant changes to $D(q)$ are visible within the large error bars: firstly, in the solution  with sodium salts there  is a moderate decrease of the effective diffusivity by 30-60\% in the large $q$-region, $q>q_{1/3}$, where the peptide is mostly unfolded. In contrast, with KCl the diffusivity seems to increase by 30-60\% in the unfolded regions.  Secondly, a drastic drop in diffusivity is observed for all salts in the central region $q_{2/3}<q<q_{1/3}$, where the peptide features one or two  $\alpha$-helical turns; the decrease is about {\it one order of magnitude} for the sodium salts NaCl and NaI.  Finally, we find a smaller decrease of the diffusivity again in the completely folded states $q<q_{2/3}$, where the diffusion drops by about  30-60\% for the KCl and NaCl salts. 
 
Thus, while no clear-cut trends in the change of the diffusivity profile with salt can be recognized, clearly the diffusivity mainly decreases with salt, along with a significant drop in the partially folded states in the intermediate $q$-region.  
Importantly, the changes of  $D(q)$ with salt are obviously not just a rescaling of $D(q)$ of the salt-free system as if the action would just stem from a nonspecific change due to a different bulk viscosity. In particular, the very few cases for which a viscosity argument applies could be for NaCl and KCl in the low $q$-region, $q<q_{2/3}$, or for NaCl and NaI in the high $q$-region, $q>q_{1/3}$: here the diffusivity is reduced by a factor of 2, roughly the same as for the viscosity increase (see Methods). Since there is no general trend, however, we must conclude that the change of the diffusivity profile originates from a combination of bulk viscosity effects {\it and} specific cosolute binding to  the peptide.  We believe that viscosity effects may be more important in the large $q$-region than in the low $q$-region: while large scale coil rearrangements in the solvent are frequent in the former, changes along the RC are mainly governed by internal mechanisms in the latter case; here only minor configurational rearrangements like the expulsion of one water molecule or one ion, or the forming and breaking of internal hydrogen-bonds take place. 
From this perspective it is interesting to see the diffusivity increasing in the high $q$-region for KCl, while from viscosity arguments only it should decrease by a factor of 2.
 
However, given the diffusivity profiles in Fig.~7, the large increase of folding times [Figs~5~(a) and 6~(a)] for the sodium salts must thus be attributed not only to the changes in the free energy landscape but more dominantly, to the strongly reduced diffusivity in the intermediate  $q$ -region,  $q_{2/3} \lesssim q\lesssim q_{1/3}$.
The faster unfolding and the weaker dependence of unfolding vs. folding times on salt type (cf. Figs.~5~(b) and 6~(b)) seems to  arise from a cancellation effect, where the small mobility in the intermediate $q$-regions is counterbalanced by the low unfolding barriers in $F(q)$ (a comparison of the FP description for average folding and unfolding times with the raw data shown in Fig.~5 is given in the SI.)
 
What are the molecular reasons for the major changes in the diffusivity profiles in the electrolytes solutions?  Based on our structural and ion binding analysis in the previous section it is now easy to argue that the huge drop in the effective diffusivity in the sodium salts is generated by the long-lived configurations similar to those shown in Fig.~4. The long-lived character of these conformations is clearly observed in trajectory analysis and also manifested in the long binding times of cations on a  nanosecond time scale shown in Fig.~3. While the form of the free energy landscape does not care about the life time
of these states (just what fraction of time they are sampled), the long life times are clearly reflected in parts of the diffusivity profiles. Due to the stronger binding of Na$^+$ vs. K$^+$ to the  peptide oxygen atoms the effect is much smaller in the KCl solution than for the sodium salts.   
 
 \section{Summary and concluding remarks}

In summary, we have investigated the specific effects of salt at molar concentration on the $\alpha$-helical folding kinetics of a short, alanine-based and salt-bridge forming peptide by means of molecular simulations and a diffusional analysis. 
Mean folding times have been found to considerably  depend on salt type with folding times varying over one order of magnitude. 
The molecular reason is the previously observed stronger binding affinity  of Na$^+$ vs. K$^+$ ions to anionic peptide groups thereby transiently cross-linking multiple groups in the peptide. 
These binding processes increase the internal friction and induce a new, slow time scale.
Within an analysis in terms of an effective diffusivity in a one-dimensional free energy  landscape, these new time scales are expressed by a strong and salt-specific variation of the local diffusivity. A recent simulation study of a fully charged polyglutamic acid
chain in salt solution showed that segmental relaxation kinetics were significantly slowed down due to the same molecular mechanisms.~\cite{Dzubiella2010}

The picture emerges that adsorption of ions not only alter the equilibrium but also kinetic properties of protein folding by direct binding mechanisms. Whether a general relation between preferential adsorption~\cite{Parsegian2000} and changes in kinetics can be drawn may be an interesting notion for further research. Given the current insights it seems likely, however, that the change in kinetics not only depends on the amount of adsorbed ions but on the individual ion-peptide interactions.
   
As we have demonstrated, molecular simulations can provide valuable information to  understand the complex mechanisms in solvent-protein interactions and thereby protein stability and folding.
The molecular mechanism found may be of general importance to	understand cosolute effects on protein folding kinetics and shed more light onto experimentally observed cation-specific slowing down of (bio)poly-electrolyte kinetics,~\cite{Colaco2008,Gray2008} in particular, for halophilic proteins;~\cite{Bandyopadhyay2007,Madern2000} similar mechanisms may be at work in polymer melts.~\cite{Mos2000}
More experimental studies are highly desirable; in particular, the novel long-lived loop-forming configurations in the denatured/unfolded states, in which sodium or similarly strong binders are bound and immobilized by the peptide backbone, may be experimentally accessible by nuclear magnetic relaxation dispersion methods (NMRD)\cite{Denisov1996} or time-resolved FRET measurement\cite{Moglich2006} probing salt-specific peptide relaxation and kinetics.

Furthermore, the action of complex denaturants such as guanidinium and urea deserve further attention, and systematic studies on specific salt effects should follow.
The guanidinium cation, for instance, has been shown to decrease friction in neutral (GlySer)$_n$ peptides.~\cite{Buscaglia2006,Moglich2006} 
We expect also a strong influence of other specifically binding cations on anionic peptides, such as lithium, or polyvalent cations, such as Mg$^{+2}$ or Ca$^{2+}$. 
Large effects may also be anticipated by exchanging the anion that has been found to considerably alter the unfolding kinetics of a halophilic protein.~\cite{Bandyopadhyay2007,Madern2000}

\section{Acknowledgments}

The authors are grateful to Michael Hinczewski and Roland R. Netz for useful discussions, the Deutsche Forschungsgemeinschaft (DFG) for support within the Emmy-Noether-Program  (IK and JD) and the SFB 863 (YvH and JD), the CompInt graduate school for support within the Elitenetzwerk Bayern (YvH), and the Leibniz Rechenzentrum (LRZ)  M\"unchen for computing time on HLRB~II. 

\section*{Supporting Information Available:}
Details regarding the choice of a specific reference structure for the definition of the RC and concerning the fitting procedure to round-trip times employed to resolve diffusivity profiles are provided in the SI. This material is available free of charge via the Internet at http://pubs.acs.org


\begin{thebibliography}{55}
\expandafter\ifx\csname natexlab\endcsname\relax\def\natexlab#1{#1}\fi
\expandafter\ifx\csname bibnamefont\endcsname\relax
  \def\bibnamefont#1{#1}\fi
\expandafter\ifx\csname bibfnamefont\endcsname\relax
  \def\bibfnamefont#1{#1}\fi
\expandafter\ifx\csname citenamefont\endcsname\relax
  \def\citenamefont#1{#1}\fi
\expandafter\ifx\csname url\endcsname\relax
  \def\url#1{\texttt{#1}}\fi
\expandafter\ifx\csname urlprefix\endcsname\relax\def\urlprefix{URL }\fi
\providecommand{\bibinfo}[2]{#2}
\providecommand{\eprint}[2][]{\url{#2}}

\bibitem[{\citenamefont{Bryngelson and Wolynes}(1989)}]{Bryngelson1989}
\bibinfo{author}{\bibfnamefont{J.~D.} \bibnamefont{Bryngelson}}
  \bibnamefont{and} \bibinfo{author}{\bibfnamefont{P.~G.}
  \bibnamefont{Wolynes}}, \bibinfo{journal}{J. Phys. Chem.}
  \textbf{\bibinfo{volume}{93}}, \bibinfo{pages}{6902} (\bibinfo{year}{1989}).

\bibitem[{\citenamefont{Camacho and Thirumalai}(1993)}]{Camacho1993}
\bibinfo{author}{\bibfnamefont{C.~J.} \bibnamefont{Camacho}} \bibnamefont{and}
  \bibinfo{author}{\bibfnamefont{D.}~\bibnamefont{Thirumalai}},
  \bibinfo{journal}{Proc. National Acad. Sciences United States Am.}
  \textbf{\bibinfo{volume}{90}}, \bibinfo{pages}{6369} (\bibinfo{year}{1993}).

\bibitem[{\citenamefont{Socci et~al.}(1996)\citenamefont{Socci, Onuchic, and
  Wolynes}}]{Socci1996}
\bibinfo{author}{\bibfnamefont{N.~D.} \bibnamefont{Socci}},
  \bibinfo{author}{\bibfnamefont{J.~N.} \bibnamefont{Onuchic}},
  \bibnamefont{and} \bibinfo{author}{\bibfnamefont{P.~G.}
  \bibnamefont{Wolynes}}, \bibinfo{journal}{J. Chem. Phys.}
  \textbf{\bibinfo{volume}{104}}, \bibinfo{pages}{5860} (\bibinfo{year}{1996}).

\bibitem[{\citenamefont{Dill and Chan}(1997)}]{Dill1997}
\bibinfo{author}{\bibfnamefont{K.~A.} \bibnamefont{Dill}} \bibnamefont{and}
  \bibinfo{author}{\bibfnamefont{H.~S.} \bibnamefont{Chan}},
  \bibinfo{journal}{Nature Struct. Biol.} \textbf{\bibinfo{volume}{4}},
  \bibinfo{pages}{10} (\bibinfo{year}{1997}).

\bibitem[{\citenamefont{Hummer et~al.}(2000)\citenamefont{Hummer, Garcia, and
  Garde}}]{Hummer2000}
\bibinfo{author}{\bibfnamefont{G.}~\bibnamefont{Hummer}},
  \bibinfo{author}{\bibfnamefont{A.~E.} \bibnamefont{Garcia}},
  \bibnamefont{and} \bibinfo{author}{\bibfnamefont{S.}~\bibnamefont{Garde}},
  \bibinfo{journal}{Phys. Rev. Lett.} \textbf{\bibinfo{volume}{85}},
  \bibinfo{pages}{2637} (\bibinfo{year}{2000}).

\bibitem[{\citenamefont{Best and Hummer}(2006)}]{Best2006}
\bibinfo{author}{\bibfnamefont{R.~B.} \bibnamefont{Best}} \bibnamefont{and}
  \bibinfo{author}{\bibfnamefont{G.}~\bibnamefont{Hummer}},
  \bibinfo{journal}{Phys. Rev. Lett.} \textbf{\bibinfo{volume}{96}}
  (\bibinfo{year}{2006}).

\bibitem[{\citenamefont{Best and Hummer}(2010)}]{Best2010}
\bibinfo{author}{\bibfnamefont{R.~B.} \bibnamefont{Best}} \bibnamefont{and}
  \bibinfo{author}{\bibfnamefont{G.}~\bibnamefont{Hummer}},
  \bibinfo{journal}{Proc. National Acad. Sciences United States Am.}
  \textbf{\bibinfo{volume}{107}}, \bibinfo{pages}{1088} (\bibinfo{year}{2010}).

\bibitem[{\citenamefont{Hinczewski et~al.}(2010)\citenamefont{Hinczewski, von
  Hansen, Dzubiella, and Netz}}]{Hinczewski2010}
\bibinfo{author}{\bibfnamefont{M.}~\bibnamefont{Hinczewski}},
  \bibinfo{author}{\bibfnamefont{Y.}~\bibnamefont{von Hansen}},
  \bibinfo{author}{\bibfnamefont{J.}~\bibnamefont{Dzubiella}},
  \bibnamefont{and} \bibinfo{author}{\bibfnamefont{R.~R.} \bibnamefont{Netz}},
  \bibinfo{journal}{J. Chem. Phys.} \textbf{\bibinfo{volume}{132}},
  \bibinfo{pages}{245103} (\bibinfo{year}{2010}).

\bibitem[{\citenamefont{Pabit et~al.}(2004)\citenamefont{Pabit, Roder, and
  Hagen}}]{Pabit2004}
\bibinfo{author}{\bibfnamefont{S.~A.} \bibnamefont{Pabit}},
  \bibinfo{author}{\bibfnamefont{H.}~\bibnamefont{Roder}}, \bibnamefont{and}
  \bibinfo{author}{\bibfnamefont{S.~J.} \bibnamefont{Hagen}},
  \bibinfo{journal}{Biochem.} \textbf{\bibinfo{volume}{43}},
  \bibinfo{pages}{12532} (\bibinfo{year}{2004}).

\bibitem[{\citenamefont{Cellmer et~al.}(2008)\citenamefont{Cellmer, Henry,
  Hofrichter, and Eaton}}]{Cellmer2008}
\bibinfo{author}{\bibfnamefont{T.}~\bibnamefont{Cellmer}},
  \bibinfo{author}{\bibfnamefont{E.~R.} \bibnamefont{Henry}},
  \bibinfo{author}{\bibfnamefont{J.}~\bibnamefont{Hofrichter}},
  \bibnamefont{and} \bibinfo{author}{\bibfnamefont{W.~A.} \bibnamefont{Eaton}},
  \bibinfo{journal}{Proc. National Acad. Sciences United States Am.}
  \textbf{\bibinfo{volume}{105}}, \bibinfo{pages}{18320}
  (\bibinfo{year}{2008}).

\bibitem[{\citenamefont{Buscaglia et~al.}(2006)\citenamefont{Buscaglia,
  Lapidus, Eaton, and Hofrichter}}]{Buscaglia2006}
\bibinfo{author}{\bibfnamefont{M.}~\bibnamefont{Buscaglia}},
  \bibinfo{author}{\bibfnamefont{L.~J.} \bibnamefont{Lapidus}},
  \bibinfo{author}{\bibfnamefont{W.~A.} \bibnamefont{Eaton}}, \bibnamefont{and}
  \bibinfo{author}{\bibfnamefont{J.}~\bibnamefont{Hofrichter}},
  \bibinfo{journal}{Biophys. J.} \textbf{\bibinfo{volume}{91}},
  \bibinfo{pages}{276} (\bibinfo{year}{2006}).

\bibitem[{\citenamefont{Moglich et~al.}(2006)\citenamefont{Moglich, Joder, and
  Kiefhaber}}]{Moglich2006}
\bibinfo{author}{\bibfnamefont{A.}~\bibnamefont{Moglich}},
  \bibinfo{author}{\bibfnamefont{K.}~\bibnamefont{Joder}}, \bibnamefont{and}
  \bibinfo{author}{\bibfnamefont{T.}~\bibnamefont{Kiefhaber}},
  \bibinfo{journal}{Proc. National Acad. Sciences United States Am.}
  \textbf{\bibinfo{volume}{103}}, \bibinfo{pages}{12394}
  (\bibinfo{year}{2006}).

\bibitem[{\citenamefont{Baldwin}(1996)}]{Baldwin1996}
\bibinfo{author}{\bibfnamefont{R.~L.} \bibnamefont{Baldwin}},
  \bibinfo{journal}{Biophys. J.} \textbf{\bibinfo{volume}{71}},
  \bibinfo{pages}{2056} (\bibinfo{year}{1996}).

\bibitem[{\citenamefont{Vrbka et~al.}(2006)\citenamefont{Vrbka, Vondrasek,
  Jagoda-Cwiklik, Vacha, and Jungwirth}}]{Vrbka2006}
\bibinfo{author}{\bibfnamefont{L.}~\bibnamefont{Vrbka}},
  \bibinfo{author}{\bibfnamefont{J.}~\bibnamefont{Vondrasek}},
  \bibinfo{author}{\bibfnamefont{B.}~\bibnamefont{Jagoda-Cwiklik}},
  \bibinfo{author}{\bibfnamefont{R.}~\bibnamefont{Vacha}}, \bibnamefont{and}
  \bibinfo{author}{\bibfnamefont{P.}~\bibnamefont{Jungwirth}},
  \bibinfo{journal}{Proc. National Acad. Sciences United States Am.}
  \textbf{\bibinfo{volume}{103}}, \bibinfo{pages}{15440}
  (\bibinfo{year}{2006}).

\bibitem[{\citenamefont{Uejio et~al.}(2008)\citenamefont{Uejio, Schwartz,
  Duffin, Drisdell, Cohen, and Saykally}}]{Uejio2008}
\bibinfo{author}{\bibfnamefont{J.~S.} \bibnamefont{Uejio}},
  \bibinfo{author}{\bibfnamefont{C.~P.} \bibnamefont{Schwartz}},
  \bibinfo{author}{\bibfnamefont{A.~M.} \bibnamefont{Duffin}},
  \bibinfo{author}{\bibfnamefont{W.~S.} \bibnamefont{Drisdell}},
  \bibinfo{author}{\bibfnamefont{R.~C.} \bibnamefont{Cohen}}, \bibnamefont{and}
  \bibinfo{author}{\bibfnamefont{R.~J.} \bibnamefont{Saykally}},
  \bibinfo{journal}{Proc. National Acad. Sciences United States Am.}
  \textbf{\bibinfo{volume}{105}}, \bibinfo{pages}{6809} (\bibinfo{year}{2008}).

\bibitem[{\citenamefont{Aziz et~al.}(2008)\citenamefont{Aziz, Ottosson,
  Eisebitt, Eberhardt, Jagoda-Cwiklik, Vacha, Jungwirth, and
  Winter}}]{Aziz2008}
\bibinfo{author}{\bibfnamefont{E.~F.} \bibnamefont{Aziz}},
  \bibinfo{author}{\bibfnamefont{N.}~\bibnamefont{Ottosson}},
  \bibinfo{author}{\bibfnamefont{S.}~\bibnamefont{Eisebitt}},
  \bibinfo{author}{\bibfnamefont{W.}~\bibnamefont{Eberhardt}},
  \bibinfo{author}{\bibfnamefont{B.}~\bibnamefont{Jagoda-Cwiklik}},
  \bibinfo{author}{\bibfnamefont{R.}~\bibnamefont{Vacha}},
  \bibinfo{author}{\bibfnamefont{P.}~\bibnamefont{Jungwirth}},
  \bibnamefont{and} \bibinfo{author}{\bibfnamefont{B.}~\bibnamefont{Winter}},
  \bibinfo{journal}{J. Phys. Chem. B} \textbf{\bibinfo{volume}{112}},
  \bibinfo{pages}{12567} (\bibinfo{year}{2008}).

\bibitem[{\citenamefont{Fedorov et~al.}(2009)\citenamefont{Fedorov, Goodman,
  and Schumm}}]{Fedorov2009}
\bibinfo{author}{\bibfnamefont{M.~V.} \bibnamefont{Fedorov}},
  \bibinfo{author}{\bibfnamefont{J.~M.} \bibnamefont{Goodman}},
  \bibnamefont{and} \bibinfo{author}{\bibfnamefont{S.}~\bibnamefont{Schumm}},
  \bibinfo{journal}{J. Am. Chem. Soc.} \textbf{\bibinfo{volume}{131}},
  \bibinfo{pages}{10854} (\bibinfo{year}{2009}).

\bibitem[{\citenamefont{Colaco et~al.}(2008)\citenamefont{Colaco, Park, and
  Blanch}}]{Colaco2008}
\bibinfo{author}{\bibfnamefont{M.}~\bibnamefont{Colaco}},
  \bibinfo{author}{\bibfnamefont{J.}~\bibnamefont{Park}}, \bibnamefont{and}
  \bibinfo{author}{\bibfnamefont{H.}~\bibnamefont{Blanch}},
  \bibinfo{journal}{Biophys. Chem.} \textbf{\bibinfo{volume}{136}},
  \bibinfo{pages}{74} (\bibinfo{year}{2008}).

\bibitem[{\citenamefont{Madern et~al.}(2000)\citenamefont{Madern, Ebel, and
  Zaccai}}]{Madern2000}
\bibinfo{author}{\bibfnamefont{D.}~\bibnamefont{Madern}},
  \bibinfo{author}{\bibfnamefont{C.}~\bibnamefont{Ebel}}, \bibnamefont{and}
  \bibinfo{author}{\bibfnamefont{G.}~\bibnamefont{Zaccai}},
  \bibinfo{journal}{Extremophiles} \textbf{\bibinfo{volume}{4}},
  \bibinfo{pages}{91} (\bibinfo{year}{2000}).

\bibitem[{\citenamefont{Bandyopadhyay et~al.}(2007)\citenamefont{Bandyopadhyay,
  Krishnamoorthy, Padhy, and Sonawat}}]{Bandyopadhyay2007}
\bibinfo{author}{\bibfnamefont{A.~K.} \bibnamefont{Bandyopadhyay}},
  \bibinfo{author}{\bibfnamefont{G.}~\bibnamefont{Krishnamoorthy}},
  \bibinfo{author}{\bibfnamefont{L.~C.} \bibnamefont{Padhy}}, \bibnamefont{and}
  \bibinfo{author}{\bibfnamefont{H.~M.} \bibnamefont{Sonawat}},
  \bibinfo{journal}{Extremophiles} \textbf{\bibinfo{volume}{11}},
  \bibinfo{pages}{615} (\bibinfo{year}{2007}).

\bibitem[{\citenamefont{Gray and Chaires}(2008)}]{Gray2008}
\bibinfo{author}{\bibfnamefont{R.~D.} \bibnamefont{Gray}} \bibnamefont{and}
  \bibinfo{author}{\bibfnamefont{J.~B.} \bibnamefont{Chaires}},
  \bibinfo{journal}{Nucleic Acids Research} \textbf{\bibinfo{volume}{36}},
  \bibinfo{pages}{4191} (\bibinfo{year}{2008}).

\bibitem[{\citenamefont{Dumetz et~al.}(2007)\citenamefont{Dumetz,
  Snellinger-O'Brien, Kaler, and Lenhoff}}]{Dumetz2007}
\bibinfo{author}{\bibfnamefont{A.~C.} \bibnamefont{Dumetz}},
  \bibinfo{author}{\bibfnamefont{A.~M.} \bibnamefont{Snellinger-O'Brien}},
  \bibinfo{author}{\bibfnamefont{E.~W.} \bibnamefont{Kaler}}, \bibnamefont{and}
  \bibinfo{author}{\bibfnamefont{A.~M.} \bibnamefont{Lenhoff}},
  \bibinfo{journal}{Protein Science} \textbf{\bibinfo{volume}{16}},
  \bibinfo{pages}{1867} (\bibinfo{year}{2007}).

\bibitem[{\citenamefont{Dyer}(1951)}]{Dyer1951}
\bibinfo{author}{\bibfnamefont{W.~J.} \bibnamefont{Dyer}},
  \bibinfo{journal}{Food Research} \textbf{\bibinfo{volume}{16}},
  \bibinfo{pages}{522} (\bibinfo{year}{1951}).

\bibitem[{\citenamefont{Lanyi}(1974)}]{Lanyi1974}
\bibinfo{author}{\bibfnamefont{J.~K.} \bibnamefont{Lanyi}},
  \bibinfo{journal}{Bacteriological Rev.} \textbf{\bibinfo{volume}{38}},
  \bibinfo{pages}{272} (\bibinfo{year}{1974}).

\bibitem[{\citenamefont{Karplus and McCammon}(2002)}]{Karplus2002}
\bibinfo{author}{\bibfnamefont{M.}~\bibnamefont{Karplus}} \bibnamefont{and}
  \bibinfo{author}{\bibfnamefont{J.~A.} \bibnamefont{McCammon}},
  \bibinfo{journal}{Nature Struct. Biol.} \textbf{\bibinfo{volume}{9}},
  \bibinfo{pages}{646} (\bibinfo{year}{2002}).

\bibitem[{\citenamefont{Marqusee and Baldwin}(1987)}]{Marqusee1987}
\bibinfo{author}{\bibfnamefont{S.}~\bibnamefont{Marqusee}} \bibnamefont{and}
  \bibinfo{author}{\bibfnamefont{R.~L.} \bibnamefont{Baldwin}},
  \bibinfo{journal}{Proc. National Acad. Sciences United States Am.}
  \textbf{\bibinfo{volume}{84}}, \bibinfo{pages}{8898} (\bibinfo{year}{1987}).

\bibitem[{\citenamefont{Marqusee et~al.}(1989)\citenamefont{Marqusee, Robbins,
  and Baldwin}}]{Marqusee1989}
\bibinfo{author}{\bibfnamefont{S.}~\bibnamefont{Marqusee}},
  \bibinfo{author}{\bibfnamefont{V.~H.} \bibnamefont{Robbins}},
  \bibnamefont{and} \bibinfo{author}{\bibfnamefont{R.~L.}
  \bibnamefont{Baldwin}}, \bibinfo{journal}{Proc. National Acad. Sciences
  United States Am.} \textbf{\bibinfo{volume}{86}}, \bibinfo{pages}{5286}
  (\bibinfo{year}{1989}).

\bibitem[{\citenamefont{Spek et~al.}(1999)\citenamefont{Spek, Olson, Shi, and
  Kallenbach}}]{Spek1999}
\bibinfo{author}{\bibfnamefont{E.~J.} \bibnamefont{Spek}},
  \bibinfo{author}{\bibfnamefont{C.~A.} \bibnamefont{Olson}},
  \bibinfo{author}{\bibfnamefont{Z.~S.} \bibnamefont{Shi}}, \bibnamefont{and}
  \bibinfo{author}{\bibfnamefont{N.~R.} \bibnamefont{Kallenbach}},
  \bibinfo{journal}{J. Am. Chem. Soc.} \textbf{\bibinfo{volume}{121}},
  \bibinfo{pages}{5571} (\bibinfo{year}{1999}).

\bibitem[{\citenamefont{Chakrabartty et~al.}(1994)\citenamefont{Chakrabartty,
  Kortemme, and Baldwin}}]{Chakrabartty1994}
\bibinfo{author}{\bibfnamefont{A.}~\bibnamefont{Chakrabartty}},
  \bibinfo{author}{\bibfnamefont{T.}~\bibnamefont{Kortemme}}, \bibnamefont{and}
  \bibinfo{author}{\bibfnamefont{R.~L.} \bibnamefont{Baldwin}},
  \bibinfo{journal}{Protein Science} \textbf{\bibinfo{volume}{3}},
  \bibinfo{pages}{843} (\bibinfo{year}{1994}).

\bibitem[{\citenamefont{Scholtz et~al.}(1991)\citenamefont{Scholtz, York,
  Stewart, and Baldwin}}]{Scholtz1991}
\bibinfo{author}{\bibfnamefont{J.~M.} \bibnamefont{Scholtz}},
  \bibinfo{author}{\bibfnamefont{E.~J.} \bibnamefont{York}},
  \bibinfo{author}{\bibfnamefont{J.~M.} \bibnamefont{Stewart}},
  \bibnamefont{and} \bibinfo{author}{\bibfnamefont{R.~L.}
  \bibnamefont{Baldwin}}, \bibinfo{journal}{J. Am. Chem. Soc.}
  \textbf{\bibinfo{volume}{113}}, \bibinfo{pages}{5102} (\bibinfo{year}{1991}).

\bibitem[{\citenamefont{Dzubiella}(2008)}]{Dzubiella2008}
\bibinfo{author}{\bibfnamefont{J.}~\bibnamefont{Dzubiella}},
  \bibinfo{journal}{J. Am. Chem. Soc.} \textbf{\bibinfo{volume}{130}},
  \bibinfo{pages}{14000} (\bibinfo{year}{2008}).

\bibitem[{\citenamefont{Case}(2006)}]{Amber}
\bibinfo{author}{\bibfnamefont{D.~A.} \bibnamefont{Case}}
  (\bibinfo{year}{2006}), \bibinfo{note}{software AMBER9.0, University of
  California, San Francisco}.

\bibitem[{\citenamefont{Dang}(1995)}]{Dang1995}
\bibinfo{author}{\bibfnamefont{L.~X.} \bibnamefont{Dang}}, \bibinfo{journal}{J.
  Am. Chem. Soc.} \textbf{\bibinfo{volume}{117}}, \bibinfo{pages}{6954}
  (\bibinfo{year}{1995}).

\bibitem[{\citenamefont{Joung and Cheatham}(2008)}]{Joung2008}
\bibinfo{author}{\bibfnamefont{I.~S.} \bibnamefont{Joung}} \bibnamefont{and}
  \bibinfo{author}{\bibfnamefont{T.~E.} \bibnamefont{Cheatham}},
  \bibinfo{journal}{J. Phys. Chem. B} \textbf{\bibinfo{volume}{112}},
  \bibinfo{pages}{9020} (\bibinfo{year}{2008}).

\bibitem[{\citenamefont{Kalcher and Dzubiella}(2009)}]{Kalcher2009}
\bibinfo{author}{\bibfnamefont{I.}~\bibnamefont{Kalcher}} \bibnamefont{and}
  \bibinfo{author}{\bibfnamefont{J.}~\bibnamefont{Dzubiella}},
  \bibinfo{journal}{J. Chem. Phys.} \textbf{\bibinfo{volume}{130}},
  \bibinfo{pages}{134507} (\bibinfo{year}{2009}).

\bibitem[{\citenamefont{Dzubiella}(2010)}]{Dzubiella2010}
\bibinfo{author}{\bibfnamefont{J.}~\bibnamefont{Dzubiella}},
  \bibinfo{journal}{J. Phys. Chem. B} \textbf{\bibinfo{volume}{114}},
  \bibinfo{pages}{7098} (\bibinfo{year}{2010}).

\bibitem[{\citenamefont{Kabsch and Sander}(1983)}]{Kabsch1983}
\bibinfo{author}{\bibfnamefont{W.}~\bibnamefont{Kabsch}} \bibnamefont{and}
  \bibinfo{author}{\bibfnamefont{C.}~\bibnamefont{Sander}},
  \bibinfo{journal}{Biopolymers} \textbf{\bibinfo{volume}{22}},
  \bibinfo{pages}{2577} (\bibinfo{year}{1983}).

\bibitem[{\citenamefont{der Spoel et~al.}(2005)\citenamefont{der Spoel,
  Lindahl, Hess, Groenhof, Mark, and Berendsen}}]{vanderSpoel2005}
\bibinfo{author}{\bibfnamefont{D.~V.} \bibnamefont{der Spoel}},
  \bibinfo{author}{\bibfnamefont{E.}~\bibnamefont{Lindahl}},
  \bibinfo{author}{\bibfnamefont{B.}~\bibnamefont{Hess}},
  \bibinfo{author}{\bibfnamefont{G.}~\bibnamefont{Groenhof}},
  \bibinfo{author}{\bibfnamefont{A.~E.} \bibnamefont{Mark}}, \bibnamefont{and}
  \bibinfo{author}{\bibfnamefont{H.~J.~C.} \bibnamefont{Berendsen}},
  \bibinfo{journal}{J. Computational Chem.} \textbf{\bibinfo{volume}{26}},
  \bibinfo{pages}{1701} (\bibinfo{year}{2005}).

\bibitem[{\citenamefont{Lindahl et~al.}(2001)\citenamefont{Lindahl, Hess, and
  van~der Spoel}}]{Lindahl2001}
\bibinfo{author}{\bibfnamefont{E.}~\bibnamefont{Lindahl}},
  \bibinfo{author}{\bibfnamefont{B.}~\bibnamefont{Hess}}, \bibnamefont{and}
  \bibinfo{author}{\bibfnamefont{D.}~\bibnamefont{van~der Spoel}},
  \bibinfo{journal}{J. Mol. Modeling} \textbf{\bibinfo{volume}{7}},
  \bibinfo{pages}{306} (\bibinfo{year}{2001}).

\bibitem[{\citenamefont{Hess}(2002)}]{Hess2002}
\bibinfo{author}{\bibfnamefont{B.}~\bibnamefont{Hess}}, \bibinfo{journal}{J.
  Chem. Phys.} \textbf{\bibinfo{volume}{116}}, \bibinfo{pages}{209}
  (\bibinfo{year}{2002}).

\bibitem[{\citenamefont{Zwanzig}(1965)}]{Zwanzig1965}
\bibinfo{author}{\bibfnamefont{R.}~\bibnamefont{Zwanzig}},
  \bibinfo{journal}{Ann. Rev. Phys. Chem.} \textbf{\bibinfo{volume}{16}},
  \bibinfo{pages}{67} (\bibinfo{year}{1965}).

\bibitem[{\citenamefont{Chen et~al.}(2009)\citenamefont{Chen, Smit, and
  Bell}}]{Chen2009}
\bibinfo{author}{\bibfnamefont{T.}~\bibnamefont{Chen}},
  \bibinfo{author}{\bibfnamefont{B.}~\bibnamefont{Smit}}, \bibnamefont{and}
  \bibinfo{author}{\bibfnamefont{A.~T.} \bibnamefont{Bell}},
  \bibinfo{journal}{J. Chem. Phys.} \textbf{\bibinfo{volume}{131}},
  \bibinfo{pages}{246101} (\bibinfo{year}{2009}).

\bibitem[{\citenamefont{Gonzalez and Abascal}(2010)}]{Gonzalez2010}
\bibinfo{author}{\bibfnamefont{M.~A.} \bibnamefont{Gonzalez}} \bibnamefont{and}
  \bibinfo{author}{\bibfnamefont{J.~L.~F.} \bibnamefont{Abascal}},
  \bibinfo{journal}{J. Chem. Phys.} \textbf{\bibinfo{volume}{132}},
  \bibinfo{pages}{096101} (\bibinfo{year}{2010}).

\bibitem[{\citenamefont{Robinson and Stokes}(2002)}]{Robinson}
\bibinfo{author}{\bibfnamefont{R.~A.} \bibnamefont{Robinson}} \bibnamefont{and}
  \bibinfo{author}{\bibfnamefont{R.~H.} \bibnamefont{Stokes}},
  \emph{\bibinfo{title}{Electrolyte Solutions}} (\bibinfo{publisher}{Dover
  Pubn. Inc.}, \bibinfo{year}{2002}), \bibinfo{note}{2nd edition}.

\bibitem[{\citenamefont{Harris and Woolf}(2004)}]{Harris2004}
\bibinfo{author}{\bibfnamefont{K.~R.} \bibnamefont{Harris}} \bibnamefont{and}
  \bibinfo{author}{\bibfnamefont{L.~A.} \bibnamefont{Woolf}},
  \bibinfo{journal}{J. Chem. Engineering Data} \textbf{\bibinfo{volume}{49}},
  \bibinfo{pages}{1064} (\bibinfo{year}{2004}).

\bibitem[{\citenamefont{Hanggi et~al.}(1990)\citenamefont{Hanggi, Talkner, and
  Borkovec}}]{Hanggi1990}
\bibinfo{author}{\bibfnamefont{P.}~\bibnamefont{Hanggi}},
  \bibinfo{author}{\bibfnamefont{P.}~\bibnamefont{Talkner}}, \bibnamefont{and}
  \bibinfo{author}{\bibfnamefont{M.}~\bibnamefont{Borkovec}},
  \bibinfo{journal}{Rev. Modern Phys.} \textbf{\bibinfo{volume}{62}},
  \bibinfo{pages}{251} (\bibinfo{year}{1990}).

\bibitem[{\citenamefont{Metzler and Klafter}(2000)}]{Metzler2000}
\bibinfo{author}{\bibfnamefont{R.}~\bibnamefont{Metzler}} \bibnamefont{and}
  \bibinfo{author}{\bibfnamefont{J.}~\bibnamefont{Klafter}},
  \bibinfo{journal}{Chem. Phys. Lett.} \textbf{\bibinfo{volume}{321}},
  \bibinfo{pages}{238} (\bibinfo{year}{2000}).

\bibitem[{\citenamefont{Barrat and Hansen}(2003)}]{Barrat}
\bibinfo{author}{\bibfnamefont{J.-L.} \bibnamefont{Barrat}} \bibnamefont{and}
  \bibinfo{author}{\bibfnamefont{J.-P.} \bibnamefont{Hansen}},
  \emph{\bibinfo{title}{Basic Concepts for Simple and Complex Fluids}}
  (\bibinfo{publisher}{University Press}, \bibinfo{address}{Cambridge},
  \bibinfo{year}{2003}).

\bibitem[{\citenamefont{Kou and Xie}(2004)}]{Kou2004}
\bibinfo{author}{\bibfnamefont{S.~C.} \bibnamefont{Kou}} \bibnamefont{and}
  \bibinfo{author}{\bibfnamefont{X.~S.} \bibnamefont{Xie}},
  \bibinfo{journal}{Phys. Rev. Lett.} \textbf{\bibinfo{volume}{93}},
  \bibinfo{pages}{180603} (\bibinfo{year}{2004}).

\bibitem[{\citenamefont{Neusius et~al.}(2008)\citenamefont{Neusius, Daidone,
  Sokolov, and Smith}}]{Neusius2008}
\bibinfo{author}{\bibfnamefont{T.}~\bibnamefont{Neusius}},
  \bibinfo{author}{\bibfnamefont{I.}~\bibnamefont{Daidone}},
  \bibinfo{author}{\bibfnamefont{I.~M.} \bibnamefont{Sokolov}},
  \bibnamefont{and} \bibinfo{author}{\bibfnamefont{J.~C.} \bibnamefont{Smith}},
  \bibinfo{journal}{Phys. Rev. Lett.} \textbf{\bibinfo{volume}{100}},
  \bibinfo{pages}{188103} (\bibinfo{year}{2008}).

\bibitem[{\citenamefont{Sangha and Keyes}(2009)}]{Sangha2009}
\bibinfo{author}{\bibfnamefont{A.~K.} \bibnamefont{Sangha}} \bibnamefont{and}
  \bibinfo{author}{\bibfnamefont{T.}~\bibnamefont{Keyes}}, \bibinfo{journal}{J.
  Phys. Chem. B} \textbf{\bibinfo{volume}{113}}, \bibinfo{pages}{15886}
  (\bibinfo{year}{2009}).

\bibitem[{\citenamefont{Parsegian et~al.}(2000)\citenamefont{Parsegian, Rand,
  and Rau}}]{Parsegian2000}
\bibinfo{author}{\bibfnamefont{V.~A.} \bibnamefont{Parsegian}},
  \bibinfo{author}{\bibfnamefont{R.~P.} \bibnamefont{Rand}}, \bibnamefont{and}
  \bibinfo{author}{\bibfnamefont{D.~C.} \bibnamefont{Rau}},
  \bibinfo{journal}{Proc. National Acad. Sciences United States Am.}
  \textbf{\bibinfo{volume}{97}}, \bibinfo{pages}{3987} (\bibinfo{year}{2000}).

\bibitem[{\citenamefont{Mos et~al.}(2000)\citenamefont{Mos, Verkerk, Pouget,
  van Zon, Bel, de~Leeuw, and Eisenbach}}]{Mos2000}
\bibinfo{author}{\bibfnamefont{B.}~\bibnamefont{Mos}},
  \bibinfo{author}{\bibfnamefont{P.}~\bibnamefont{Verkerk}},
  \bibinfo{author}{\bibfnamefont{S.}~\bibnamefont{Pouget}},
  \bibinfo{author}{\bibfnamefont{A.}~\bibnamefont{van Zon}},
  \bibinfo{author}{\bibfnamefont{G.~J.} \bibnamefont{Bel}},
  \bibinfo{author}{\bibfnamefont{S.~W.} \bibnamefont{de~Leeuw}},
  \bibnamefont{and} \bibinfo{author}{\bibfnamefont{C.~D.}
  \bibnamefont{Eisenbach}}, \bibinfo{journal}{J. Chem. Phys.}
  \textbf{\bibinfo{volume}{113}}, \bibinfo{pages}{4} (\bibinfo{year}{2000}).

\bibitem[{\citenamefont{Denisov et~al.}(1996)\citenamefont{Denisov, Peters,
  Horlein, and Halle}}]{Denisov1996}
\bibinfo{author}{\bibfnamefont{V.~P.} \bibnamefont{Denisov}},
  \bibinfo{author}{\bibfnamefont{J.}~\bibnamefont{Peters}},
  \bibinfo{author}{\bibfnamefont{H.~D.} \bibnamefont{Horlein}},
  \bibnamefont{and} \bibinfo{author}{\bibfnamefont{B.}~\bibnamefont{Halle}},
  \bibinfo{journal}{Nature Struct. Biol.} \textbf{\bibinfo{volume}{3}},
  \bibinfo{pages}{505} (\bibinfo{year}{1996}).

\bibitem[{\citenamefont{Humphrey et~al.}(1996)\citenamefont{Humphrey, Dalke,
  and Schulten}}]{Humphrey1996}
\bibinfo{author}{\bibfnamefont{W.}~\bibnamefont{Humphrey}},
  \bibinfo{author}{\bibfnamefont{A.}~\bibnamefont{Dalke}}, \bibnamefont{and}
  \bibinfo{author}{\bibfnamefont{K.}~\bibnamefont{Schulten}},
  \bibinfo{journal}{J. Mol. Graphics} \textbf{\bibinfo{volume}{14}},
  \bibinfo{pages}{33} (\bibinfo{year}{1996}).

\end{thebibliography}

\clearpage 

Fig.~1: Time series data for reaction coordinate $q$ from the MD simulations in explicit salt water with varying salt type.
Data in black shows the full 20~ps resolution, while data in red is smoothed over time windows of 2~ns. (figure not enclosed)
 
\vspace{1cm}

Fig.~2: 
(a)~Free energies $F(q)$ for the EK peptide in different salt solutions; the profiles are shifted vertically for 
better comparison. (b)~Average $\alpha$-helicity of the peptide resolved by q. The snapshots illustrate the backbone structure of partially folded/unfolded states corresponding to values of the helicity indicated by arrows. Simulation snapshots are visualized using VMD \cite{Humphrey1996}. 

\vspace{1cm}

Fig:~3: Radial distribution function of cations around peptide oxygen atoms from the side chain (sc) carboxylates or backbone (bb) carbonyls. 
Inset: residence time distribution for the cations in the first solvation shell of the peptide oxygen atoms. The distribution for K$^+$ can 
be fitted by a single exponential with a time constant of $\tau=$~50~ps, while the one for Na$^+$ obeys a stretched exponential with $\tau=$~50~ps and stretching exponent $\beta=0.55$. 

\vspace{1cm}

Fig.~4: Simulation snapshots of backbone configurations, in which a single Na$^+$-ion (blue sphere) is trapped by multiple oxygen 
binding sites (red). These snapshots are dynamically selected such that the shown configurations existed for longer 
than 2~ns. In (a) and (b) no side chains are shown to better illustrate the binding to the backbone. In (c) and (d) all side chains
are shown. Here, also glutamic acid side chains are involved in binding the cation.  (figure not enclosed)

\vspace{1cm}

Fig.~5: (a)~Mean first passage times $\tau_\text{f}$ for folding from state $q \rightarrow q_1$. (b)~Mean first passage times $\tau_\text{uf}$ 
for unfolding from state $q\rightarrow q_3$.

\vspace{1cm}

Fig.~6:~(a) Mean first passage times $\tau_\text{f}$ for folding to state $q \rightarrow q_2$. (b)~Mean first passage times $\tau_\text{uf}$ 
for unfolding to state $q\rightarrow q_2$. 

\vspace{1cm}

Fig.~7:~(a) Free energy landscapes (same as in Fig.~2~(a)). (b)~Diffusivity profiles $D(q)$ for all investigated systems  resolved by three regions $q<q_{2/3}$, $q_{2/3}\leq q\leq q_{1/3}$, and $q\geq_{1/3}$, which correspond to the fully helical, two-turn, and one-turn as well as unfolded states, respectively. Symbols including error bars result from fits to the ensemble of round-trip times $\tau_\text{RT}(q,q_\text{t})$ for different targets $q_\text{t}$, while fitting results to the average round-trip time curve $\bar{\tau}_\text{RT}$ (Eq.~\ref{eq:round-trip_avg}) are displayed as solid lines.

\vspace{1cm}

Fig.~8: Table of contents (TOC) figure.  (figure not enclosed)

%

\newpage 

{\large Figure 2}
\begin{figure}[htp]
 \includegraphics[width=14cm,angle=0]{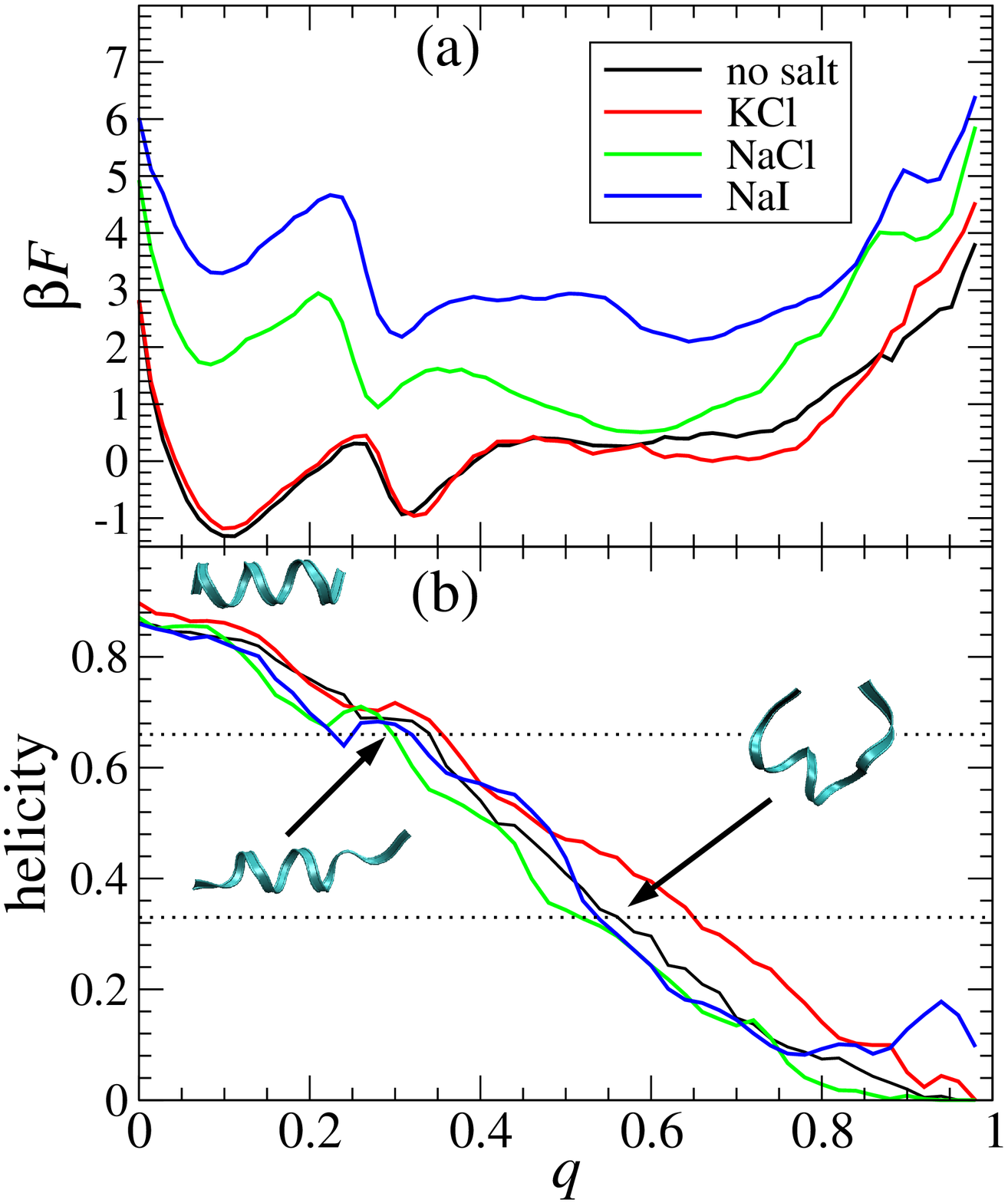}
\end{figure}

\newpage 

{\large Figure 3}
\vspace{2cm}
\begin{figure}[htp]
 \includegraphics[width=14cm,angle=0]{fig3.eps}
\end{figure}

%

\newpage 

{\large Figure 5}
\vspace{2cm}
\begin{figure}[h]
 \includegraphics[width=14cm,angle=0]{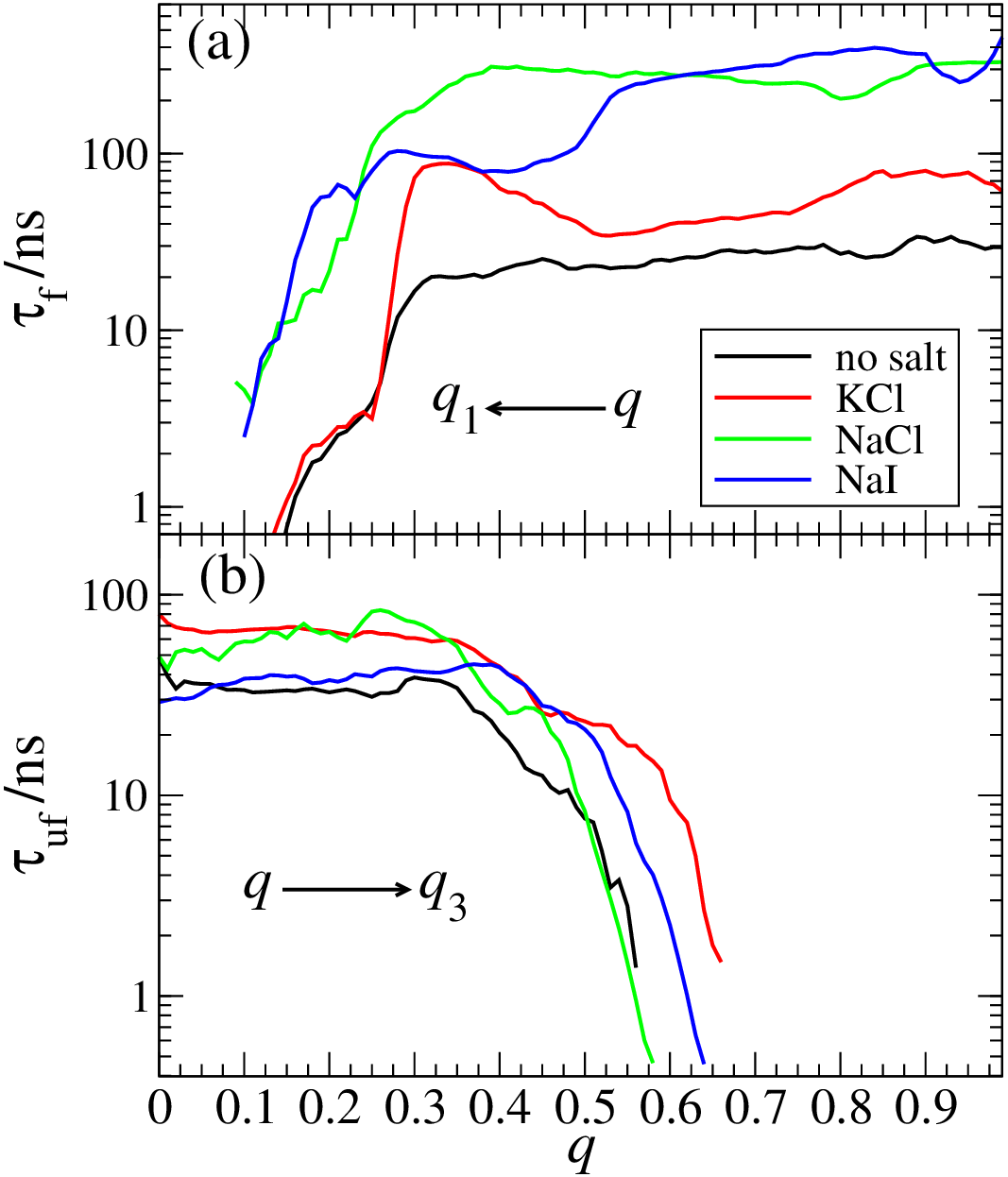}
\end{figure}

\newpage 

{\large Figure 6}
\vspace{2cm}
\begin{figure}[htp]
 \includegraphics[width=14cm,angle=0]{fig6.eps}
\end{figure}

\newpage 

{\large Figure 7}
\vspace{2cm}
\begin{figure}[htp]
 \includegraphics[width=14cm,angle=0]{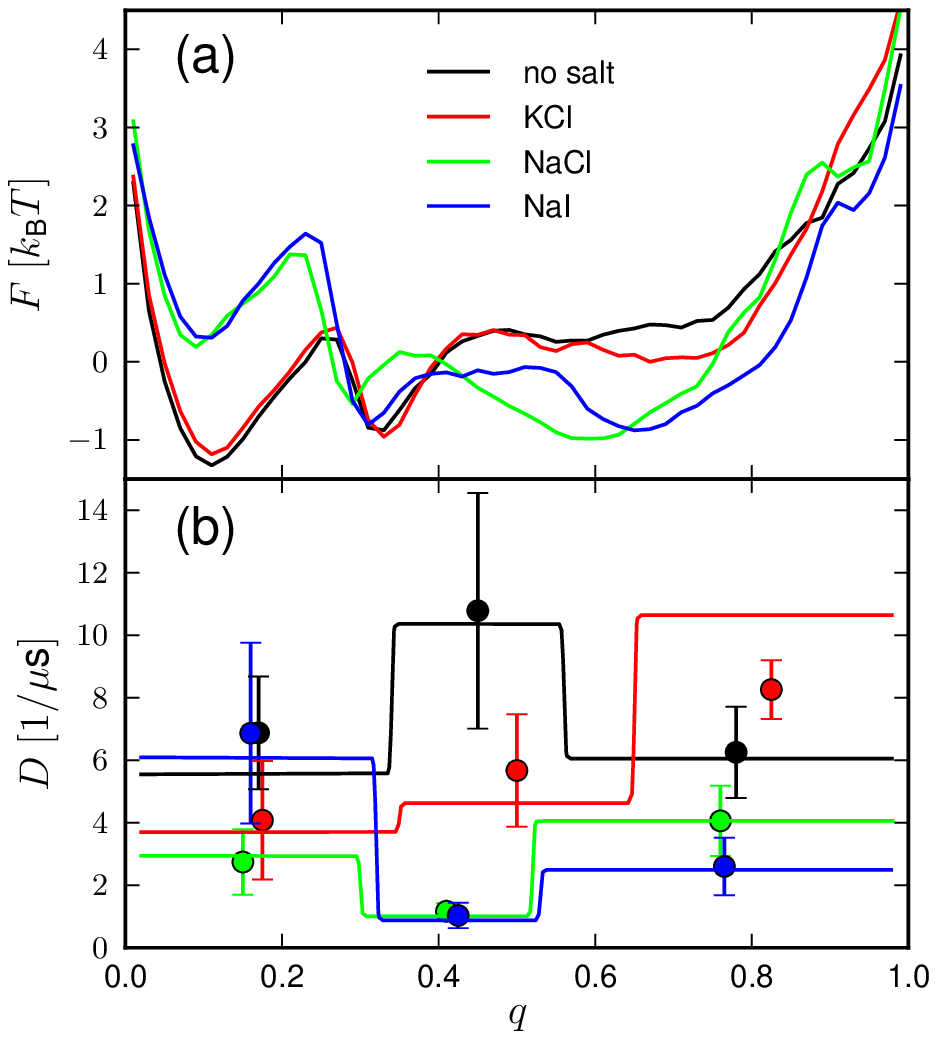}
\end{figure}

%
%

\end{document}